\documentclass[11pt]{article}
\usepackage{amsmath}    
\usepackage{amssymb}
\usepackage{graphicx}
\usepackage{amsfonts}   
\usepackage{latexsym}   
\usepackage{slashed}
\usepackage{yfonts} 
\usepackage{verbatim}
\usepackage{wrapfig}
\usepackage{mathrsfs}
\usepackage{cancel}
\usepackage[super]{nth}
\usepackage[normalem]{ulem}
\usepackage{color}

\global\newcount\itemno \global\itemno=0

\def\itemaut#1{\global\advance\itemno by1\noindent\item{\the\itemno.}#1}

\newif{\ifShowComputations}           

\newcommand{\be}{\begin{equation}}
\newcommand{\ee}{\end{equation}}
\newcommand{\bes}{\begin{equation*}}
\newcommand{\ees}{\end{equation*}}
\newcommand{\bea}{\begin{eqnarray}}
\newcommand{\eea}{\end{eqnarray}}
\newcommand{\beas}{\begin{eqnarray*}}
\newcommand{\eeas}{\end{eqnarray*}}

\def\({\left(}
\def\){\right)}
\def\[{\left[}
\def\]{\right]}

\def\frac#1#2{{#1 \over #2}}

\newcommand{\g}{\gamma}

\renewcommand{\l}{\lambda}

\newcommand{\m}{\mu}

\renewcommand{\S}{\Sigma}

\newcommand{\Tr}{{\rm Tr \,}}

\newcommand{\lsim}{\,\raise.3ex\hbox{$<$\kern-.75em\lower1ex\hbox{$\sim$}}\,}
\newcommand{\gsim}{\,\raise.3ex\hbox{$>$\kern-.75em\lower1ex\hbox{$\sim$}}\,}

\def\II{\relax{I\kern-.10em I}}

\usepackage{amssymb}

\numberwithin{equation}{section}
\usepackage[colorlinks,linkcolor=blue,citecolor=blue]{hyperref}

\usepackage{multirow}
\usepackage{mciteplus}

\newcommand{\K}{\mathcal{K}}
\renewcommand{\g}{g_{\mu\nu}}
\renewcommand{\S}{{\rm Sec.}}

\newcommand{\HMG}{HMG}

\begin{document}

\begin{titlepage}
\begin{flushright}MIT-CTP/4578 \end{flushright}
\vskip 1in

\begin{center}

{\Large{\bf{Hawking-Page transition in holographic massive gravity}}}

\vskip 0.5in Allan Adams,$^{a}$ Daniel A. Roberts,$^{a}$ and Omid Saremi$^{b}$
\vskip 0.2in {\it $^{a}$ Center for Theoretical Physics \\  Massachusetts Institute of Technology\\ Cambridge, MA  02139, USA}
\vskip 0.2in {\it $^{b}$ Department of Physics and Astronomy \\ University of British Columbia \\ Vancouver, B.C., V6T 1W9, Canada}
\end{center}
\vskip 0.5in

\begin{abstract}
\noindent
We study the Hawking-Page transition in a holographic model of field theories with momentum dissipation.  
We find that the deconfinement temperature strictly decreases as momentum dissipation is increased. 
For sufficiently strong momentum dissipation, the critical temperature goes to zero, indicating a 
zero-temperature deconfinement transition in the dual field theory.  

\end{abstract}

\end{titlepage}

\baselineskip=17.63pt

%
%
%
\section{Introduction}

Spatial inhomogeneity has profound consequences for the dynamics of real physical systems.  For single particle systems the detailed consequences of inhomogeneity are well understood, with periodicity leading to band structure and disorder leading to Anderson localization \cite{Anderson:1958vr}.  On macroscopic scales over which the system looks effectively homogeneous, these effects are simply encoded in bulk material properties such as efficient momentum dissipation and a finite DC resistivity.  In strongly interacting many-body systems, however, the consequences of disorder remain far more mysterious.  When does disorder lead to localization?  Can disorder interfere with confinement or other many-body effects?  What are the macroscopic legacies of microscopic disorder in a strongly interacting many-body system?

Holography provides a powerful set of tools with which to attack these questions.  
Holographically, momentum conservation in a homogeneous QFT follows from spatial diffeomorphism invariance in the gravitational bulk.\footnote{More precisely, conservation of the QFT's stress-energy tensor follows from the asymptotic limit of the mixed radial-boundary components of the bulk Einstein equations.}  The holographic dual of an effective QFT with momentum dissipation should thus violate translational diffeomorphism invariance in the bulk.  This can be arranged by, for example, turning on spatially inhomogeneous boundary sources dual to bulk scalar fields.  Linearizing in this inhomogeneous perturbation gives a Stuckelberg-like mass term for certain spatial components of the bulk metric.  The effective field theory of such a spatially massive graviton thus provides a homogeneous effective holographic description of the underlying inhomogeneous system valid on wavelengths long compared to the microscopic scale of inhomogeneity.\footnote{{For other approaches to modeling inhomogeneity in holographic systems, see for example \cite{Hartnoll:2007ih,Hartnoll:2008hs,Adams:2011rj,Adams:2012yi,Hartnoll:2012rj,Anantua:2012nj,Lucas:2014zea,Hartnoll:2014cua,AdamsSaremiUP,Andrade:2013gsa}.}}

Such a holographic effective field theory has been worked out by a series of authors  \cite{Vegh:2013sk, Blake:2013bqa, Davison:2013txa,Blake:2014yla,Amoretti:2014zha} based on the formalism of \cite{Hinterbichler:2011tt, deRham:2014zqa, Hassan:2011vm}. In this effective theory, known as holographic massive gravity (\HMG), the nonconservation of momentum is manifest in a modified conservation law for the stress tensor
\be
\partial_\mu T^{\mu i}=-\tau^{-1}T^{t i}\,, \label{noncontinuity}
\ee
which is valid for sufficiently high temperatures.
Here $T^{\mu \nu}$ is the momentum current, $i$ is a spatial index, and $\mu$ is a space-time index.  Since $T^{\mu i}$ is the momentum current, \eqref{noncontinuity} expresses the nonconservation of momentum, with $\tau^{-1}$ measuring the system's momentum relaxation rate.  
In \HMG, the near-equilibrium relaxation rate $\tau^{-1}$ is \cite{Blake:2013bqa,Davison:2013jba}
\be
\tau^{-1}=\frac{s}{4\pi} \frac{m_{h}^2}{u+P},
\ee
where $s$, $u$, $P$, are the entropy density, energy density, and pressure, and $m_{h}^2$ is the mass squared of the graviton (evaluated at the horizon). 
The essential feature is that the relaxation time is inversely proportional to the mass---the graviton mass directly encodes the rate of momentum dissipation in the dual QFT.  It is then straightforward to use this description to model effectively homogeneous systems with manifestly finite DC conductivity \cite{Vegh:2013sk, Blake:2013bqa, Davison:2013txa}.

This model has been shown to agree in linearization to the backreaction to a background space-time lattice \cite{Blake:2013owa}\ and has been motivated via a Goldstone mechanism for spontaneous breaking of translational symmetry by an inhomogeneous matter background.  However, it has not been derived as an effective action, and indeed the precise regime of validity and how to derive the macroscopic parameters governing dissipation from the microscopic details of the spatial inhomogeneity remain unknown.  For the moment, we will simply bracket such concerns and study the physics of the fully nonlinear holographic massive gravity as a model of a momentum-dissipating theory at strong coupling.  

The present note explores the thermodynamic phase structure of these theories.  As we shall see, turning on a holographic mass parameter decreases the critical temperature of the Hawking-Page transition, driving it to zero temperature at a finite value set by the AdS radius.  These results lead to some simple conjectures about the phase structure of strongly coupled field theories with strong momentum dissipation---in particular, that increasing the efficiency of momentum dissipation (by, say, increasing the strength of quenched disorder in a strongly coupled spin glass) strictly decreases the deconfinement temperature,  leading to a zero-temperature disorder-driven deconfinement transition at finite values of the momentum dissipation rate.

%
%
%

\section{The Hawking Page transition in AdS}\label{ads-review}

We begin by reviewing the thermodynamics of $3+1$-dimensional Anti-de Sitter space (AdS$_{4}$) in Einstein gravity plus cosmological constant \cite{Hawking:1982dh}.  We will work in global coordinates with metric
\be
ds^2 = \frac{L^2}{r^2} \left(\frac{dr^2}{f(r)} - f(r)dt^2 +L^2 d\Omega^2 \right), \label{global-metric}
\ee 
where $d\Omega^{2}$ is the round metric on the 2-sphere. The boundary is a 2-sphere whose radius is given by the AdS radius $L$ and occurs at $r=0$. For different choices of the {emblackening factor}, $f(r)$, the above metric can describe both thermal AdS, with
\be
f(r)=1+\frac{r^2}{L^2},
\ee
and Schwarzschild AdS, with
\be
f(r)=1+\frac{r^2}{L^2}-Mr^3. \label{global-bh}
\ee
The Schwarzschild-AdS black hole solution has a horizon at the smallest $r$ such that $f(r)=0$, denoted $r_h$.

The temperature $T$ of the thermal AdS solution is put in by hand by taking the imaginary time direction to be compact with period $1/T$. The temperature of the black hole solution is given by
\be
T=-{f'(r_h)\over 4\pi } = {1\over 4 \pi r_h} \( 3+{r_{h}^{2}\over L^{2}} \)\,, 
\ee
and comes from demanding that the Euclidean continuation be regular at the horizon. Solving for $r_h$, we find
\be
r_h=2 \pi  L^2 T  \pm \sqrt{(2 \pi L^2 T)^2 -3L^2}\,, \label{two-sols}
\ee
which is real if
\be
(2 \pi L^2 T)^2 -3L^2 \ge 0.
\ee
This shows that there are no black hole solutions in global AdS below temperature\footnote{For any  $T>T_0$, there are two black hole solutions given by the positive and negative branches of \eqref{two-sols}. The larger black hole has positive specific heat while the smaller black hole has negative specific heat.  In the canonical ensemble in which we work, holding the temperature fixed, this small black hole is thus unstable. In the microcanonical ensemble, holding the energy fixed, these small black holes are dynamically stable, coming to dynamical equilibrium with their Hawking radiation \cite{Horowitz:1999uv}.}
\be\label{TO}
T_0=\frac{\sqrt{3}}{2 \pi  L}.
\ee

This has an important consequence for the stability of a thermal gas in AdS.  In flat space, any sufficiently large thermal gas will collapse under its own gravity and form a black hole.  For temperatures sufficiently large compared to the AdS radius, then, we should expect a uniform thermal gas in global AdS to collapse to form a black hole.  However, at temperatures below $T_{0}$, the thermal gas cannot form an AdS black hole and is thus stable.  This suggests that there should be a thermal phase transition in AdS at some temperature $T_{HP} \ge T_{0}$, with the thermal gas dominating at temperatures below $T_{HP}$ and the black hole dominating above $T_{HP}$.  This is the Hawking-Page transition.

Computing the critical Hawking-Page temperature, $T_{HP}$, is a straightforward exercise in black hole thermodynamics.  Working in the canonical ensemble with fixed temperature $T$, the dominant solution is the one with the minimal free energy.  For these space-times, the free energy is given by their on-shell Euclidean action,\footnote{Without renormalization, this action is ill-defined.  A proper calculation requires the addition of boundary terms such as the Gibbons-Hawking-York term, with which to properly renormalize the on-shell action.   
For the case at hand, these boundary counterterms completely cancel when computing the difference of free energies and so can safely be neglected.  In general, this is not the case.}
\be
S_E[g]=-\frac{1}{16\pi G_N}\int d^4x\sqrt{g}~\Big\{R + \frac{6}{L^2}\Big\},
\ee
after analytic continuation $t \to - i\tau$, with the temperature determined by the period of the Euclidean time coordinate, $\tau \sim \tau + 1/T$. 
The Hawking-Page transition occurs at the point where the actions are equal. Evaluating the actions and subtracting one from the other gives
\be
S_E[f_{BH}(r)]-S_E[f_{AdS}(r)]= \frac{  L^4}{4 G_N r_h^3 T} \Big(
\frac{r_h^2}{L^2} - 1
\Big),
\ee
which vanishes when $r_h=L$.  
Using \eqref{two-sols}, this gives the Hawking-Page temperature
\be
T_{HP}=\frac{1}{\pi  L},
\ee
which is greater than $T_0$ by a factor of $2/\sqrt{3}  \sim 1.15$.

%
%
%

\section{The Hawking Page transition in holographic massive gravity}\label{massive-section}

Our goal is to identify the Hawking-Page transition in holographic massive gravity as introduced by Vegh in \cite{Vegh:2013sk}, following the earlier work developing the dRGT theory of massive gravity \cite{deRham:2010ik, deRham:2010kj}, with action\footnote{{We discuss the consequences of including other powers of $\Tr\K$ in Appendix~\ref{generalizations-tr}}.  In the literature, the coefficient $-{\mu^2\over 2}$ is usually written as $\beta$.}
\begin{align}
S=\frac{1}{16\pi G_N} \int d^4x \sqrt{-g}\Big\{R+\frac{6}{L^2}- \frac{\mu^2}{2} \big[ (\Tr\K)^2 - \Tr\K^2\big]  \Big\}. \label{full-action} 
\end{align}
Here $\sqrt{-g}R$ is the usual Einstein-Hilbert action in $D=3+1$ dimensions, $L$ is the AdS radius, $G_N$ is the $3+1$-dimensional Newton's constant, and the parameter $\mu$, with dimensions of mass, parametrizes the violation of diffeomorphism invariance.
$\K$ is a matrix defined by
\be
\K^\mu_{\ \rho} \K^\rho_{\ \nu}=g^{\mu\sigma} f_{\sigma\nu}.
\ee
$g^{\mu\sigma}$ is the inverse metric, and $f_{\sigma\nu}$ is a symmetric 2-tensor called the {\em reference metric}. 

Diffeomorphism invariance is broken by nonzero values of the reference metric, $f_{\sigma\nu}$.  Since we are interested in preserving homogeneity and isotropy on the spatial sphere, as well as general $r$-$t$ diffeomorphism invariance, the natural choice for  $f_{\sigma\nu}$ is the $SU(2)$-invariant $f_{\theta\theta}=L^2$ and $f_{\phi\phi}=L^2\sin^2 \theta$ (where $\theta$ and $\phi$ are the angular coordinates parametrizing the boundary 2-sphere) with all the other components vanishing. This gives
\be
\K^\theta_{\ \theta}=\K^\phi_{\ \phi}=\frac{r}{L},
\ee
and zero for the other components. The resulting action is still invariant under general diffeomorphisms that mix $r$ and $t$, as well as rigid rotations of the spatial sphere, but does not respect arbitrary reparametrizations involving $\theta$ or $\phi$.

\subsection{Graviton mass}

An important constraint on the parameter values comes from requiring the graviton mass to be positive.  Since $\mu^2$ controls the breaking of spatial diffeomorphism invariance, and thus the mass of corresponding modes of the graviton, bounds on the (sign of the) mass of the graviton become bounds on the allowed values of $\mu^2$.  In \cite{Blake:2013bqa} the (radially dependent) mass of the graviton in asymptotically Poincare-AdS backgrounds was reported as
\be
m^2(r)=\mu^2. \label{graviton-mass}
\ee
This mass was derived in planar coordinates for the $t-x$ or $t-y$ components by considering fluctuations around the massive gravity background, plugging in the zeroth-order solution, and identifying the linear terms with $\mu^2$ dependence (including contributions from the Einstein-Hilbert action). 
Repeating this analysis in asymptotically global AdS solutions, we find for the $h_{t\theta}$ fluctuation the equation of motion
\be
h_{t\theta}''+\frac{2 h_{t\theta}'}{r}-\frac{2 h_{t\theta}}{r^2} +\frac{2 h_{t\theta}}{L^2 f(r)}= \frac{ \mu^2  h_{t\theta}}{f(r)}.
\ee
Identifying the terms on the right-hand side as
\be
\frac{m^2(r)}{f(r)} h_{t\theta}, \label{mass-norm}
\ee
gives the same result as in the Poincare patch, \eqref{graviton-mass}.

Requiring $m^2(r) \ge 0$ for local stability thus implies that $\mu^2>0$.  Precisely how one should define the graviton mass turns out to be a subtle question.  The na\"ive definition and bound above fit nicely with several other constraints on the allowed values of $\mu^2$ (such as one coming from consistency of the black hole solution below, or the holographic interpretation of $\mu^2$ as controlling the rate of momentum dissipation in the dual theory \cite{Davison:2013jba}). So, in the absence of a better definition we use this as a good working definition and bound.

\subsection{The black hole solution}

The equations of motion resulting from \eqref{full-action} admit spherically symmetric asymptotically AdS solutions of the same form as in \eqref{global-metric}, but with modified emblackening function,
\be
f(r)=1 + \frac{r^2}{L^2}\Big[1-\frac{\mu^{2}L^{2}}{2}\Big] - M r^3, \label{f(M)}
\ee
where $M$ is a constant proportional to the energy density of the black hole.

%
%
\ifdefined\ShowCalculations

As usual, there is a horizon at any radius $r=r_{h}$ such that $f(r_{h})=0$. Trading $M$ for $r_{h}$, $f(r)$ becomes
\be
f(r)=1+ \frac{r^2}{L^2}\Big[1-\frac{\mu^{2}L^{2}}{2}\Big]  -\Big(\frac{1}{r_h^3}+ \frac{1}{r_h L^2 }\Big[1-\frac{\mu^{2}L^{2}}{2}\Big]  \Big)r^3, \label{f(rh)}
\ee
which clearly satisfies $f(r_h)=0$.  Requiring regularity of the Euclidean continuation gives the black hole temperature
\be
T=-\frac{f'(r_h)}{4\pi}=\frac{1}{4\pi r_h}\Big(3  + \frac{r_h^2}{L^2}\Big[1-\frac{\mu^{2}L^{2}}{2}\Big] \Big)\,. \label{temperature}
\ee
Solving for $r_h$ in terms of  $T$ and $\mu^2$ gives
\be
r_h=
\Big[1-\frac{\mu^{2}L^{2}}{2}\Big]^{-1} \Big(2 \pi  L^2 T
 \pm \frac{1}{2}
 \sqrt{\left(4 \pi  L^2 T\right)^2-12 L^2 \Big[1-\frac{\mu^{2}L^{2}}{2}\Big]} \Big)      ,\label{rh-eq}
\ee
which reproduces \eqref{two-sols} as $\mu^2\to0$. The horizon radius $r_h$ is real when
\be
\left(4 \pi  L^2 T\right)^2-12 L^2 \Big[1-\frac{\mu^{2}L^{2}}{2}\Big] \ge 0. \label{reality-rh}
\ee
If \eqref{reality-rh} cannot be satisfied for a particular value of $T$, then no black hole solutions exists at that temperature. Rearranging, we find
\be\label{TOM}
T_0=\frac{\sqrt{3}}{2\pi L}\sqrt{1-\frac{\mu^{2}L^{2}}{2}}.
\ee
selecting the positive branch to get a positive temperature. Below this temperature, holographically massive AdS does not support black hole solutions.

{Note that \eqref{f(M)} and  \eqref{TOM} takes the same form as \eqref{global-bh} and  \eqref{TO} up to replacing the AdS radius $L$ with an effective AdS radius,
\be
\tilde{L}\equiv \frac{L}{\sqrt{1-\frac{\mu^{2}L^{2}}{2}}}.
\ee
We avoid this substitution in what follows so as to keep the $\mu$ dependence transparent. However, this has an important implication to which we shall return below.}
%
%
%
%
\else
{Note that \eqref{f(M)} takes the same form as \eqref{global-bh} up to replacing the AdS radius $L$ with an effective AdS radius,
\be
\tilde{L}\equiv \frac{L}{\sqrt{1-\frac{\mu^{2}L^{2}}{2}}}.
\ee
Repeating the analysis above then gives the same results up to the substitution $L\to \tilde{L}$, giving the black hole temperature as
\be
T=\frac{1}{4\pi r_h}\Big(3  + \frac{r_h^2}{L^2}\Big[1-\frac{\mu^{2}L^{2}}{2}\Big] \Big)\,. \label{temperature}
\ee
and allowing us to conclude that no black holes exist below the temperature 
\be\label{TOM}
T_0(\m)=\frac{\sqrt{3}}{2\pi L}\sqrt{1-\frac{\mu^{2}L^{2}}{2}}.
\ee
}
\fi
%
%

\subsection{Thermal solution} \label{thermal-solution-section}

Before we can discuss the analog of the Hawking-Page transition, we need to identify the analog of thermal AdS. Unfortunately, global AdS is no longer a solution in massive gravity. Instead, we look for solutions without a horizon which we can interpret as the massive analog of global AdS, with temperature fixed by the periodicity of the Euclidean time coordinate.

We can do this by taking $r_h\to\infty$ or, equivalently seen from \eqref{f(M)}, by taking the mass  to zero. In order that the space-time not have a horizon, $f(r)$ must not have a zero. Solving $1 + [1-\mu^{2}L^{2}/2]~r_h^2/L^2=0 $ for $r_h$  gives us
\be
r_h=
\frac{L}{ \sqrt{-1+\frac{\mu^{2}L^{2}}{2}}}.\label{thermal-horizon}
\ee
If the discriminant is negative then there is no horizon, leading to the no-horizon condition
\be
\mu^2<\frac{2}{L^2}. \label{no-horizon-condition}
\ee
The resulting space-time is the unique spherically symmetric horizon-free solution to the massive-gravity equations of motion, providing the natural $\mu^2$-deformed generalization of global AdS.   
Note that there is thus no (spherically symmetric) horizon-free solution for $\m^{2}>{2\over L^{2}}$.  For greater values of $\m^{2}$, we will inescapably have a horizon.\footnote{Although they have a horizon, these $\mu^2 > 2/L^2$ black holes have a negative heat capacity and are thermodynamically unstable
\be
T\frac{dS(T)}{dT}=-\frac{ \Big[1-\frac{\mu^{2}L^{2}}{2}\Big]^2}{2 \pi G_N  T^2}.
\ee
Thus, they are not seen by the canonical ensemble and will play no role in our phase diagram.}

Unlike global AdS, the Ricci scalar in this solution is not constant,  
\be
R=-\frac{12}{L^2} +\frac{ \mu^2 }{L^2}r^2\,,
\ee
revealing the presence of a naked singularity at  $r\to\infty$.  For our purposes, however, this singularity is relatively innocuous.  For example, since $\propto \sqrt{-g}R$ vanishes as $1/r^2$, the singularity does not contribute to the free energy (the on-shell Euclidean action).  Meanwhile, the singularity is a finite proper distance (and time) from any observer in the bulk, effectively cutting off the IR portion of the geometry and reflecting a gap in the dual system.   Finally, unlike other nakedly singular solutions such as Lifshitz metrics, adding a small temperature does {\em not} introduce a horizon: no black hole solutions exist below temperature $T_{0}(\m)$. {It does make the computation of low-energy correlation functions somewhat involved; for thermodynamic purposes, however, it appears we can proceed without resolving the singularity.}

\subsection{Hawking-Page transition}

We are now equipped to study the fate of the Hawking-Page transition in our model of holographic massive gravity.  As in the case of global AdS, the difference between the Euclidean actions of the two solutions computes their relative free energy in the canonical ensemble. The solution with the lowest free energy is thermodynamically dominant; a Hawking-Page transition occurs when they are equal. 

The Euclidean action is given by
\be
S_E=-\frac{1}{16\pi G_N}\int d^4x\sqrt{g}~\Big\{R + \frac{6}{L^2}  - \frac{\mu^2}{2} \big[ (\Tr\K)^2 - \Tr\K^2\big]  \Big\}.
\ee
As before, $\K^\mu_{\ \rho} \K^\rho_{\ \nu}=g^{\mu\sigma} f_{\sigma\nu}$ and $f_{\mu\nu}=diag(0,0,L^2,L^2 \sin^2 \theta)$.\footnote{Contrast this to \cite{Vegh:2013sk} or \cite{Blake:2013bqa} where in Poincare coordinates, they use $f_{\mu\nu}=diag(0,0,1,1)$.}  
%
%
%
\ifdefined\ShowCalculations
Unlike the case of global AdS, it is crucial to properly renormalize the bulk action for general HMG Lagrangians. 
Following \cite{Blake:2013bqa}, we find that we need the counterterm action,
\be
S_{ct}= \frac{1}{16\pi G_N}\int d^3x\sqrt{\gamma}~\Big\{-2\Theta +\frac{4}{L} \Big(1+ \frac{\epsilon^2}{2L^2} + 
A+ B\frac{\epsilon}{L}   + C\frac{\epsilon^3}{L^3} + D\frac{\epsilon^4}{L^4}\Big)\Big\} \label{counter-terms}
\ee
where $\gamma$ is the determinant of the induced metric on the boundary cutoff surface at $r=\epsilon$, and $\Theta=\gamma^{\mu\nu}\nabla_{\mu}n_{\nu}$ is the extrinsic curvature, giving the standard Gibbons-Hawking-York boundary term. Requiring regularity as the cutoff is removed fixes 
\be
A=0, \qquad B=0, \qquad C=-\frac{\mu^2  L^2}{4}.
\ee
These terms are sufficient to render the action finite to take $\epsilon\to0$. Importantly, the $D$ term is finite in the limit, and it conspires with leftover $r_h$ or $M$ independent pieces to add an additional constant to the free energy. 

Computing the action and taking the difference, we find
%
%
%
%
\else
Upon carefully renormalizing and repeating the Hawking-Page analysis, we find that the difference in the free energies between black hole and thermal state is given by
\fi
%
%
%
\be
S_E[f(r); M]-S_E[f(r); 0] = \frac{  L^4}{4 G_N r_h^3 T} ~
\Big(
\frac{r_h^2}{L^2}~\Big[1-\frac{\mu^{2}L^{2}}{2}\Big] -1
\Big), \label{action-diff}
\ee
which vanishes when $r_h={L\over\sqrt{1-\mu^{2}L^{2}/2}}=\tilde{L}$. Using \eqref{temperature}, we find 
\be
T_{HP}(\mu^2)=\frac{\sqrt{1-\frac{\mu^{2}L^{2}}{2}}}{\pi  L}\,,
\ee
which is always greater than $T_0$, as in the usual Hawking-Page transition.

\subsection{Phase diagram}

The critical temperature for massive gravity in global coordinates is
\be
T_{HP}(\mu^2)=\frac{1}{\pi \tilde{L}} = \frac{1}{\pi  L}\sqrt{1-\frac{\mu^{2}L^{2}}{2}} \,,
\ee
When $\mu^2=0$, this reproduce the Hawking-Page transition for global AdS as expected. 
As noted above, $\tilde{L}$  plays the role of an effective AdS length in that, while the radius of the spatial sphere remains $L$, the thermodynamic relations are identical to pure AdS with $L\to\tilde{L}$.  As we send $\mu^2$ to $\mu^2_* \equiv 2/L^{2}$,  this effective AdS radius {\em diverges} and the critical temperature {\em vanishes}. Thermodynamically, it is as if our system effectively lived on a very small patch of the sphere, recovering the results of the decompactified Poincare patch.

\begin{figure}[t]
\begin{center}
\includegraphics[scale=0.8]{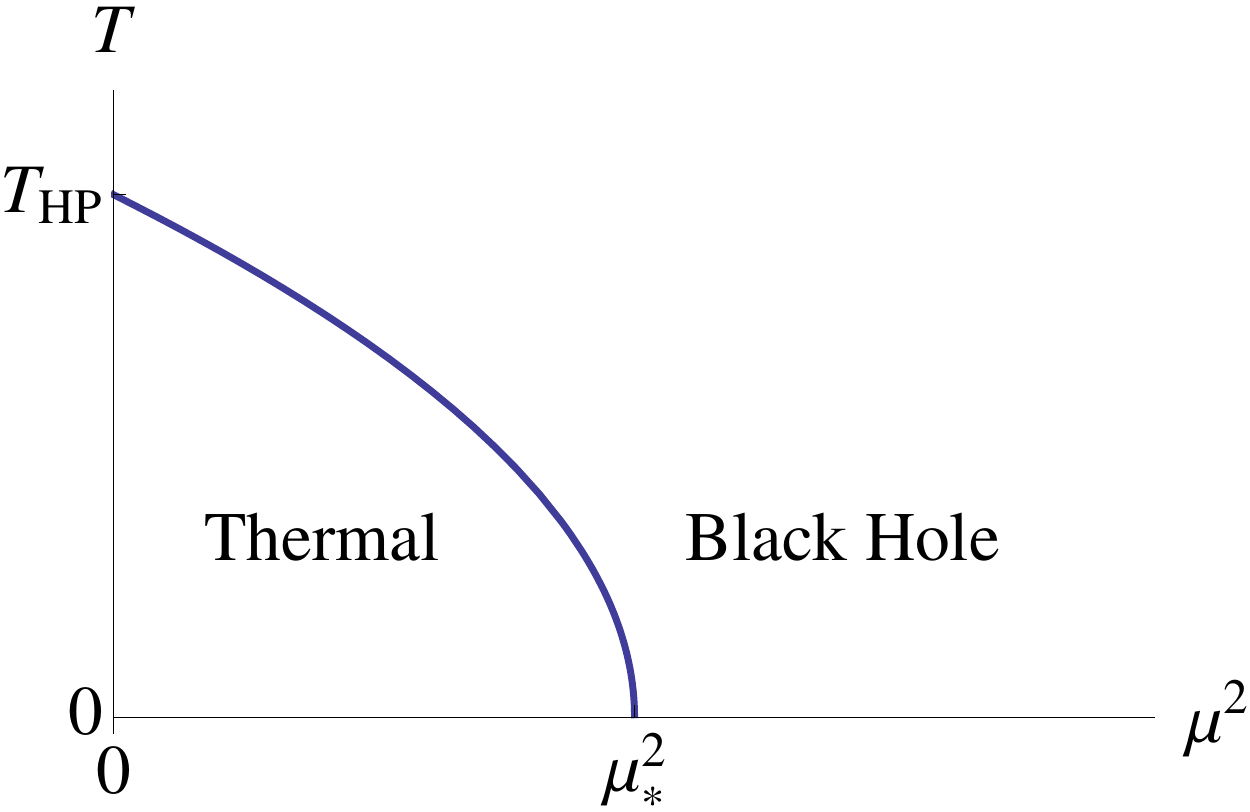}
\end{center}
\caption{Phase diagram for holographic massive gravity showing the Hawking-Page temperature, $T_{HP}(\mu^2)$, vs.\! the mass parameter, $\mu^2$. The region below the purple line is a confined phase in which the thermal solution dominates the ensemble, while the region above the line is a deconfined phase in which the black hole solution dominates. The line of first-order phase transitions terminates in a second-order phase transition at zero temperature and $\mu^2 = \mu^2_*=2/L^2$.}
\label{phase-diagram}
\end{figure}

Figure~\ref{phase-diagram} shows the resulting phase diagram for holographic massive gravity. At $\m^{2}=0$ we recover the usual result in AdS: a first-order phase transition in which both energy and entropy jump from $O(1)$ to $O(N^2)$ as the temperature rises past $T_{HP}$ due to the appearance of a black hole with finite horizon area. For $0<\mu^2<2/L^{2}$, the temperature of this first-order transition decreases but the qualitative features are the same. At $\mu^2=\mu^2_*$, however, this line of first-order phase transitions terminates in a continuous phase transition---both the entropy and the energy are continuous. The energies must be equal (and, in fact, are vanishing) since the system is at zero temperature and the free energies are equal. The entropy is continuous because the black hole horizon to the right of  $\mu^2_*$ approaches $r_h=\infty$, and thus has vanishing area.

%
%
%
\section{Discussion}\label{thoughts-puzzles}

The results above suggest that varying $\mu^2$ leads to a zero-temperature deconfinement phase transition.  From the point of view of the field theory, this might seem odd.  By construction, $\mu^2$ controls the rate of momentum dissipation in the field theory, as can be seen 
from the form of the stress-tensor Ward identities and their violation (cf. \cite{Davison:2013jba}) and can be attributed to quenched disorder \cite{Vegh:2013sk,Davison:2013txa}.  Why is momentum dissipation inducing a deconfinement transition?

Consider first the $\mu^2=0$ system.  At $T=0$, the system confines with a confinement scale set by the AdS radius $L$. 
On scales longer than the confinement scale, charges are completely screened, so that the only propagating modes in the deep IR are a set of $O(1)$ singlets.  On scales shorter than the confinement scale, however, all $N^{2}$ degrees of freedom are accessible and propagate. This is a heuristic way to understand the Hawking-Page transition in AdS---deconfinement occurs when the thermal scale $\beta$ is less than the confinement scale.

Turning on momentum dissipation changes the story.  Intuitively, if modes are allowed to shed momentum, they will be less efficient at screening charges.  This suggests that allowing momentum dissipation should increase  the confinement length (equivalently, decrease the gap in the confining regime).  This is precisely what we find above.  Indeed, from simple dimensional analysis, $\mu^2$ defines a new length scale, $\l_{\mu^2}=1/\mu$, which we can roughly understand as the distance a particle with $O(1)$ momentum will travel before dissipating its momentum.  For small $\mu^2$, this scale is very long---much longer than the confinement scale---and the effect is to moderately diminish the efficiency of confinement and gently decrease the gap, leaving the free energy $O(1)$.  Effectively, over the length scale $\l_{\mu^2}$, the only propagating modes are a set of $O(1)$ singlets, so it is natural that the change in the free energy remains $O(1)$.   This is consistent with what we noted before, that the effect of $\mu^2$ is to shift the effective AdS radius for the critical temperatures from $L$ to a strictly larger $\tilde{L}$.  As $\mu^2$ is further increased, the effect grows, with the gap shrinking further and further.

When momentum dissipation grows strong, something curious happens---the length scale $\l_{\mu^2}$ grows smaller than the clean confinement scale (indeed at $\mu^2=\mu^2_*$, the effective AdS radius blows up; the system effectively becomes infinitely large).  Physically, excitations with $O(1)$ momentum cannot propagate beyond the confinement scale.  For all practical purposes, they live in a system with an infinite confinement length.  For $\mu^2>\mu^2_{*}$, the full $N^{2}$ microscopic degrees of freedom are effectively localized on a scale below the ambient confinement scale.  They are deconfined.

A prediction of this simple heuristic is that increasing the efficiency of momentum dissipation beyond some critical threshold should do two things: first, it should deconfine the system, with the entropy scaling as $N^{2}$; second, the free energy density should depend not on the size of the system, $L$, but on the dissipation scale $\l_{\mu^2}$, as one would expect of a system whose effective size is $\l_{\mu^2}$.  The first is confirmed by the dominance of the black hole solution down to $T=0$ for $\mu^2>\mu^2_{*}$, generalizing the usual finite-temperature deconfinement transition.  The second is apparent in the thermodynamics of the $T=0$, $\mu^2>\mu^{2}_{*}$ black hole, whose dual energy density is given by,\footnote{{If the fact that the energy density is negative bothers you, remember we can shift the free energy by a constant $U_0$ by using the $D$ counterterm in our holographic renormalization \cite{Blake:2013bqa}.}} 
\be
U=-\frac{L^{4}}{6 \sqrt{6} G_N} \( \mu^{2} -\m^{2}_{*} \)^{3/2}.
\ee
Dividing by the area of the $2$-sphere ($4\pi L^2$) and substituting for $G_N$, we find the energy density
\be
u=-\frac{N^2 \(   \mu^{2} -\m^{2}_{*}  \)^{3/2}}{24 \sqrt{6} \pi },
\ee
which, for $\mu^2 \gg \m^{2}_{*}$ becomes
\be
u=-N^2\frac{\mu ^{3}  }{24 \sqrt{6} \pi  } \sim \frac{N^2}{\l_{\mu^2}^3}.
\ee
We thus confirm that the free energy density scales as $N^{2}$ and depends on the dissipation scale, $\l_{\m}$, rather than the AdS scale, $L$, as suggested by our heuristic interpretation above.

\subsection*{Puzzles}

Thus far we have taken HMG at face value as a long-distance model of a QFT with momentum dissipation.  To the degree that \HMG\ is well defined, the above results follow.  However, we remain puzzled by a number of things about holographic massive gravity and the metrics it generates.

First and foremost, there is no derivation of \HMG\ as an effective theory obtained by integrating out a spatially inhomogeneous matter background in pure Einstein gravity.  To be sure, there have been several heuristic and semiquantitative checks of this picture \cite{Blake:2013bqa,Davison:2013jba,Blake:2013owa}, but all of these treat the source of translation noninvariance perturbatively and linearize the gravitational response.  In this paper, by contrast, we have taken the full nonlinear \HMG\ theory seriously.  Whether this is physically reasonable remains an open question.  To that end, it would be interesting to compare the fully nonlinear massive gravity theory to the holographic model of massless scalars explored in \cite{Andrade:2013gsa}. There, the authors realize momentum relaxation via spatially dependent constant-gradient sources. These theories agree at the linear level, so it would be interesting to see if they are actually the same theory.\footnote{We thank Richard Davison for this suggestion.}

Meanwhile, the naked singularity in the thermal solution described in \S~\ref{thermal-solution-section}\ remains a source of some confusion.  On the one hand, such IR singularities are to be expected in systems exhibiting strong disorder, as explored in \cite{Adams:2011rj,Adams:2012yi,Hartnoll:2014cua}. Moreover, the fact that these singularities live at finite proper distance from points in the bulk {indicates that there is a gap in the dual field theory}.  On the other hand, this naked singularity makes the study of e.g. transport in the deep IR (below the Hawking-Page temperature) somewhat subtle.  Note that this puzzle arises even for perturbative values of $\mu^2$.  We hope to return to these puzzles in the future.


\section*{Acknowledgments}
We  thank Mike Blake, Richard Davison, Sean Hartnoll, Sung-Sik Lee, Hong Liu, Alex Maloney, Moshe Rozali, Douglas Stanford, David Tong, and David Vegh for helpful discussions.  
A.A. thanks the Aspen Center for Physics for hospitality during the completion of this work.
The work of A.A. and D.R. is supported in part by the U.S. Department of Energy under cooperative research agreement DE-FG02-05ER41360.
The work of D.R. is also supported in part by the NDSEG Program and by the Fannie and John Hertz Foundation.
The work of O.S. is supported by the Natural Sciences and Engineering Research Council of Canada (NSERC).

\appendix

%

\ifdefined\ShowCalculations

\section{Constructing and Solving the EOMs of HMG}\label{eom-derive}

We will vary the inverse metric to get the equation of motion from the action\footnote{In this section, we will follow the conventions of the literature, setting $\mu^2=-2\beta$.}
\begin{align}
S&=S_1+S_2, \\
S_1 &= \frac{1}{16\pi G_N} \int d^4x \sqrt{-g}\Big\{R+\frac{6}{L^2} \Big\}, \\
S_2 &= \frac{1}{16\pi G_N} \int d^4x \sqrt{-g}\Big\{ \alpha \Tr\K + \beta \big[ (\Tr\K)^2 - \Tr\K^2\big]  \Big\}. \\ \label{fullest-action} 
\end{align}
The variation of the $S_1$ term simply gives Einstein's equations with a negative cosmological constant,
\be
\frac{1}{\sqrt{-g}} \frac{\delta S_1}{\delta g^{\mu\nu}} = R_{\mu\nu} - \frac{1}{2}R \g - \frac{3}{L^2}\g .
\ee
The variation of the $S_2$ term gives the modification due to the graviton mass term
\be
\frac{1}{\sqrt{-g}} \frac{\delta S_2}{\delta g^{\mu\nu}} = X_{\mu\nu},
\ee
which we will derive below.
\subsection*{$\K$ variations}
In matrix form (which will be more convenient in the following),
\begin{align}
&\K^2=g^{-1}f,\\
&\K=\sqrt{g^{-1}f}.
\end{align}
We will need various variations of $\K$, which we will derive here. We will work using the matrix form of all expressions, and our two biggest ``tricks" will be cyclically of the trace to ease differentiation commutation ambiguities and  insertions of $g^{-1}g=I$.

\subsubsection*{$\delta\Tr\K$ terms}
\begin{align}
\delta \Tr\K &= \Tr \delta (g^{-1}f)^{\frac{1}{2}}, \notag \\
 &= \frac{1}{2}\Tr (g^{-1}f)^{-\frac{1}{2}} \delta (g^{-1}f), \notag \\
&= \frac{1}{2}\Tr \K^{-1} \delta g^{-1}f, \notag \\
&= \frac{1}{2}\Tr \K^{-1} \delta g^{-1}  g (g^{-1} f), \notag \\
&=  \frac{1}{2}\Tr \left(\delta g^{-1}  g K \right).
\end{align}
In component form this reads
\begin{align}
\delta \K^\mu_{\ \mu} &= \frac{1}{2}  \delta g^{\mu \rho} g_{\rho \sigma} \K^\sigma_{\ \mu}, \notag \\
&= \frac{1}{2}  \delta g^{\mu \rho}  \K_{\rho\mu},
\end{align}
where
\be
\K_{\rho\mu} = g_{\rho \sigma} \K^\sigma_{\ \mu}.
\ee
That is, we will lower and raise indices on $\K$ using the metric $\g$.

We will also need the variation of powers of $\Tr\K$,
\begin{align}
\delta (\Tr\K)^p &= p(\Tr\K)^{p-1}\delta\Tr\K, \notag \\
&=  \frac{p}{2} (\Tr\K)^{p-1} \Tr \left(\delta g^{-1}  g K \right). \label{TrKp}
\end{align}
For the action $S_2$, we will only require $p=2$, but later we will consider adding other terms $\sim (\Tr\K)^p$ for $p>2$.

\subsubsection*{$\delta\Tr\K^q$ terms}

We will also have to vary terms like $\sim \Tr\K^q$ for $q=2$, but we will record the general result here as well.
\begin{align}
\delta \Tr\K^q &= q \Tr\K^{q-1} \delta\K, \notag \\
&=  q \Tr \K^{q-1} \left( \frac{1}{2}\delta g^{-1}  g K\right), \notag \\
&= \frac{q}{2} \Tr \left(\delta g^{-1} g \K^{q} \right). \label{TrKq}
\end{align}

\subsection*{$\delta S_2$ variation}

The final thing we need to compute the variation is to remember that 
\be
\delta \sqrt{-g} = -\frac{1}{2}\sqrt{-g}\g\delta g^{\mu\nu}.
\ee
Putting it all together, we find
\be
X_{\mu\nu} = \frac{\alpha}{2}\Big( \K_{\mu\nu} - (\Tr\K)\g\Big) + \beta \Big( (\Tr\K) \K_{\mu\nu} -(\K^2)_{\mu\nu}  + \frac{1}{2}\g (\Tr \K^2 - (\Tr\K)^2)  \Big),
\ee
with the EOM
\be
R_{\mu\nu} - \frac{1}{2}R \g - \frac{3}{L^2}\g + X_{\mu\nu} =0. \label{eom}
\ee

\subsection*{Black hole solution}
Since we are looking for black brane solutions with a negative cosmological constant, we assume a metric ansatz for spherically symmetric static space times
\be
ds^2 = \frac{L^2}{r^2} \left( \frac{dr^2}{f(r)} - f(r)dt^2 +d\Omega^2 \right), \label{appendix-metric}
\ee
where the boundary lies at $r=0$. Plugging the ansatz into \eqref{eom}, the $t-t$ equation gives a differential equation for $f(r)$
\be
0=-\frac{f(r) \left(L^2 \left(r f'(r)-3 f(r)\right)+L^2 \left(\alpha  L r+\beta  r^2+3\right)+r^2\right)}{L^2 r^2}, \label{eomtt}
\ee
which is easily solved to give
\be
f(r)=1+\frac{r^2}{L^2}+\frac{L\alpha r}{2} + \beta r^2 -M r^3.
\ee
This solution for the planar black hole was first found in  \cite{Vegh:2013sk}.

\fi
%
%

\section{Generalization and extensions}\label{generalizations-tr}
{\HMG\ can be generalized in several ways. Here we repeat our analysis for several such modifications.}

\subsection{Higher dimensions}\label{general-D}
The original Hawking-Page transition was found for AdS$_4$ \cite{Hawking:1982dh}. In  \cite{Witten:1998zw} it was subsequently shown that this phase transition persists to all $D$, with a $D$-dependent formula for the Hawking-Page temperature $T_{HP}$. Here we show that the holographically massive Hawking-Page transition also persists to all $D$ and also show that the temperature below which black holes cannot exist in global AdS ($T_0$) has a similar $D$-dependent generalization.

We will work global coordinates with the metric ansatz, 
\be
ds^2 = \frac{L^2}{r^2} \left(\frac{dr^2}{f(r)} - f(r)dt^2 +L^2 d\Omega^2_{D-2} \right), \label{D-metric}
\ee 
where $d\Omega^{2}_{D-2}$ is the round metric on the $(D-2)$-sphere. 

We will consider the following modified action
\begin{align}
S&=S_1+S_2, \label{full-action-D-dimensions}\\
S_1&=\frac{1}{16\pi G_N} \int d^Dx \sqrt{-g}~\Big\{R+\frac{(D-1) (D-2)}{L^2}\Big\}, \\
S_2&=\frac{1}{16\pi G_N} \int d^Dx \sqrt{-g}~\Big\{-\frac{\m^{2}}{2} [ (\Tr\K)^2 - \Tr\K^2]\Big\},  \label{S2}
\end{align}
where $S_1$ is the usual Einstein-Hilbert action in $D$ dimensions, $L$ is the AdS radius, $G_N$ is now the $D$-dimensional Newton's constant, and $S_2$ is the massive gravity addition that breaks diffeomorphism invariance.  As before, the  parameter $\m^{2}$ has dimensions of mass squared and $\K$ is a matrix defined by $\K^\mu_{\ \rho} \K^\rho_{\ \nu}=g^{\mu\sigma} f_{\sigma\nu}.$
Here, $g^{\mu\sigma}$ is the inverse metric, and $f_{\sigma\nu}$ is a symmetric 2-tensor which is given by $diag(0,0,h_{\sigma\nu})$, where $h_{\mu\nu}dx^\mu dx^\nu=L^2 d\Omega^2_{D-2}$ is proportional to the metric on the $(D-2)$-sphere. This breaks diffeomorphism invariance in a rotationally invariant way on the spatial $(D-2)$-sphere while at the same time preserving $r$-$t$ diffeomorphisms. If we wanted, we could also add to action $S_2$ additional terms polynomial in $\Tr\K$. The effect of a term $\Tr\K^p$  is to add a term to the emblackening factor $\sim r^p$ (see \S~\ref{BOne}). For now, we will set these terms to zero.

Substituting ansatz \eqref{D-metric} into the equations of motion resulting from \eqref{full-action-D-dimensions}, we find the unique spherically symmetric solution,
\be
f(r)=1 + \frac{r^2}{L^2}[1-\frac{\mu^{2}L^{2}}{2}]  - M r^{D-1}, \label{f(M,D)}
\ee
where again, $M$ is proportional to the energy density of the black hole.  The temperature of the black hole is given by
\be
T =  \frac{ (D-3) \Big[1-\frac{\mu^{2}L^{2}}{2}\Big]r_h^2  +   \left(D-1\right)L^2    }{4 \pi  L^2 r_h},\label{temp-D}
\ee
which recovers the correct expression for AdS$_{D}$ black holes when $\mu^2=0$.
As before, we will consider the thermal solution to be the $M=0$ case of this metric.  As expected, this solution has a naked singularity, with the free energy again insensitive to the singularity since $\sqrt{-g}R$ vanishes as $1/r^{D-2}$ at infinity.

Solving for $r_h$ and demanding that the horizon radius is real, we find
\be
T_0 =  \frac{\sqrt{(D-3) (D-1) [1-\frac{\mu^{2}L^{2}}{2}]}}{2 \pi  L}.
\ee
Below this temperature black holes cannot exist. For $\mu^2=0$, we see that in AdS$_D$ we have the following generalization
\be
T_0 =  \frac{\sqrt{(D-3) (D-1) }}{2 \pi  L}.
\ee

Similarly, we can compute the difference of the Euclidean actions of the black hole and thermal solutions to find a Hawking-Page transition. Taking care to properly handle the boundary terms, we find that the difference between the actions is zero at the same horizon radius independent of dimension
\be
r_h={L\over\sqrt{1-\frac{\mu^{2}L^{2}}{2}}}.
\ee
Plugging this into \eqref{temp-D}, we find the Hawking-Page temperature to be
\be
T_{HP}=\frac{(D-2)\sqrt{1-\frac{\mu^{2}L^{2}}{2}}}{2\pi  L},
\ee
which correctly recovers the correct expression for AdS$_{D}$ when $\mu^2=0$.  Finally, we note that $T_{HP}>T_0$ as expected for consistency.

\subsection{Finite density}

We can also consider the holographically massive extension of Reissner-Nordstrom black holes. If we consider solutions of holographically massive Einstein-Maxwell theory with emblackening factor 
\be
f(r)=1+ \frac{r^2}{L^2}\Big[1-\frac{\mu^{2}L^{2}}{2}\Big] -Mr^3+\frac{\phi^2}{r_h^2 \gamma^2}r^4, 
\ee
with
\be
M=\frac{1}{r_h^3}+ \frac{1}{r_h L^2 }\Big[1-\frac{\mu^{2}L^{2}}{2}\Big]+\frac{\phi^2}{r_h \gamma^2},
\ee
and
\be
\gamma^2=\frac{e^2L^2}{4\pi G_N},
\ee
then we find, for instance, a critical temperature $T_0$ that depends on the chemical potential as
\be
T_0 = \frac{\sqrt{3}}{2}\frac{\sqrt{1-\mu^{2}L^{2}/2 - \phi^2 L^2/\gamma^2}}{\pi  L},
\ee
so $\phi^2$ plays the same role in the critical behavior as does $\mu^2$. The comments in \cite{Chamblin:1999tk} suggest that extremal RN black holes are in a confined phase; here, the presence of the massive graviton allows the $T=0$ state to be deconfined for large enough $\mu^2$.

\subsection{More general reference metrics $f_{\mu\nu}$}
We might also imagine a more general reference metric that does not respect diffeomorphism invariance in the $r$-direction.  

For example, if we want to preserve rotational invariance on the sphere and general time reparametrization invariance, one natural ansatz is $f_{\mu\nu} = diag(0,1,L^2,L^2 \sin^2 \theta)$. 
The resulting $G^{tt}$ equation evaluated on the spherically symmetric metric ansatz \eqref{global-metric} unfortunately does not admit a simple solution.
The complication is that the emblackening factor $f(r)$ now appears in $\K$: while $\K^\theta_{\ \theta}=\K^\phi_{\ \phi}=r/L$ as before, $\K^r_{\ r}= rf(r)^{\frac{1}{2}} /L$. This leads to a nonlinear ODE for $f(r)$.

Alternatively, as suggested in \cite{Blake:2013bqa}, one can also break diffeomorphism invariance in the $r$-direction by considering a different generalization of $f_{\mu\nu}$, with $\sin^2 \theta f_{\theta\theta}=f_{\phi\phi}=g(r)$, and all other components to be zero. Ultimately this leads to an ability to add arbitrary polynomial terms in $r$ to the emblackening factor (similar to the generalizations below in \ref{BOne}).

\subsection{Higher order trace terms}\label{BOne}
{In the bulk of this note, we introduced only quadratic terms to the graviton action.  More generally we could add an arbitrary polynomial,}
\be
S_3=\frac{1}{16\pi G_N}  \int d^4x \sqrt{-g} ~   \sum \alpha_p \left( \Tr \K \right)^p  ,
\ee
where the parameters $\alpha_p$ all have the dimensions of mass squared. These terms (excepting $p=1$) are normally not included since they induce the Boulware-Deser ghost \cite{Boulware:1973my}, which the action \eqref{S2} is finely tuned to eliminate \cite{deRham:2010kj}. 
However, any reference metric $f_{\mu\nu}$ that preserves time reparametrization invariance will not induce the ghost \cite{Blake:2013bqa}.

We compute the variation of $S_3$ to be
\be
\frac{1}{\sqrt{-g}} \frac{\delta S_3}{\delta g^{\mu\nu}} = Y_{\mu\nu} = \sum_{p=3} \frac{\alpha_p}{2} \Big(p(\Tr \K)^{p-1}\K_{\mu\nu} - \left(\Tr \K \right)^p\g \Big)
\ee
The $t$-$t$ component of the equation of motion evaluated on the spherically symmetric metric ansatz now includes an additional term
\be
Y_{00}=\sum_p 2^{p-1}\alpha_p f(r)\Big(\frac{L^2}{r^2}\Big)\bigg(\frac{r}{L}\bigg)^p.
\ee
This new differential equation for $f(r)$ can be solved to give an emblackening factor with higher order $r$ terms
\be
f(r)=1+\frac{r^2}{L^2}+\frac{L\alpha r}{2} + \beta r^2 -M r^3 - \sum_p \frac{2^{p-1}L^2 \alpha_p }{p-3} \bigg(\frac{r}{L}\bigg)^p.
\ee
The $p=3$ case, which includes a term $\sim r^3 \log r$, is special.

\mciteSetMidEndSepPunct{}{\ifmciteBstWouldAddEndPunct.\else\fi}{\relax}
\bibliographystyle{utphys}
\bibliography{phase}{}

\end{document}